\documentclass[12pt]{article}
\usepackage{graphicx,amsmath,amssymb,color,bm}

\begin{document}

\centerline {\Large\textbf {Electrically tunable plasma excitations in AA-stacking}}
\centerline {\Large\textbf {multilayer graphene}}


\centerline{Ming-Fa Lin, Ying-Chih Chuang, Jhao-Ying Wu$^{*}$}
\centerline{Department of Physics, National Cheng Kung University,
Tainan, Taiwan 701}

\vskip0.6 truecm

\noindent

We use a tight-binding model and the random-phase approximation to study the Coulomb excitations in simple-hexagonal-stacking multilayer graphene and discuss the field effects. The calculation results include the energy bands, the response functions, and the plasmon dispersions. A perpendicular electric field is predicted to induce significant charge transfer and thus capable of manipulating the energy, intensity, and the number of plasmon modes. This could be further validated by inelastic light scattering or electron-energy-loss spectroscopy.

\vskip0.6 truecm

$\mathit{PACS}$: 73.21.Ac, 73.22.Lp, 73.22.Pr

\newpage

\bigskip

\centerline {\textbf {I. INTRODUCTION}}%

\bigskip

\bigskip

Multilayer graphene (MLG), a stack of graphene layers, draws a great deal of attention recently. The main reason is that its energy bands strongly depends on the stacking sequence and the number of layers,$^{1-5}$ and can be easily modulated by external fields.$^{6-9}$ Those features directly reflect on their transport properties and suggest the possible applications in electronic devices.$^{10-11}$ The three typical stacking modes are Bernal, rhombohedral, and simple-hexagonal stacking, i.e., ABA, ABC,$^{12}$ and AAA.$^{13}$ The naturally occurring MLGs are usually in ABA or ABC types and their properties have been widely studied in many respects.$^{1-11,14-16}$ Their energy bands are anisotropic and mostly parabolic-like owing to the relative displacements between the adjacent layers. A perpendicular electric field can induce a band overlap in odd-layer ABA (mirror symmetry) but a band gap in even-layer ABA and in all ABC MLG. On the other hand, AA-stacking is often encountered in growing graphene.$^{17}$ It is found recently that bilayer graphene tends to exist in this form and is very hard to distinguish from a single layer graphene in scanning tunneling microscopy.$^{18}$ The highly symmetrical configuration normal to the layers causes the energy bands of the AA MLG to resemble those of a collection of independent monolayer graphenes. The dispersion near the Fermi level is linear and isotropic, which implies a high-efficiency carrier transport, and better than in a monolayer graphene, the density of free carriers can be tuned by a perpendicular electric field for a fixed chemical potential. So far most theoretical studies on AA MLG focus on its electronic properties$^{19,20}$ or single particle excitations.$^{21,22}$ To understand other physical properties requires more researches.

Collective Coulomb excitation is a powerful theoretical and experimental probe to explore the fundamental physical properties of materials. A low-frequency plasmon hardly exists in undoped monolayer graphene due to the vanishing density of states at the Fermi level.$^{23}$ However, it is expected to survive in AA bilayer graphene,$^{24}$ which has intrigued us to further investigate the plasmon effects in AA MLG. We utilize the tight-binding model and the random-phase approximation (RPA) $^{25}$ to calculate the dielectric function, which is cast into a matrix form to include both the interlayer atomic interactions and interlayer Coulomb interactions.$^{26}$ The intensity of the various intralayer and interlayer charge polarizations can be modulated by a perpendicular electric field, which directly reflects on the collective excitations. Since the tight-binding model incorporates the whole $\pi$-band structure, the calculated excitation spectrum is reliable over a wide energy region, and the exact band structure allows the field strength to be tuned in a wide range. The method can deal with different number of layers and various stacking sequences (ABA and ABC shown in supporting information) and may be employed in many studies of MLG. The predicted results could be further verified by optical experiments$^{27-29}$ or electron-energy-loss spectroscopy (EELS).$^{30-32}$


\bigskip
\bigskip
\centerline {\textbf {II. METHODS}}%
\bigskip
\bigskip

A graphene sheet is a hexagonal lattice of carbon atoms. It is
composed of two sublattices \emph{A} and \emph{B}, and the C-C bond
length is \emph{b}=1.42 {\AA}. The geometric structure of an AA-stacking
\emph{N}-layer graphene is that all carbon atoms on every sheet are
stacked directly over those of the adjacent sheets with an interlayer
distance of \emph{c}=3.35 {\AA}. There are 2\emph{N} carbon
atoms ($A_{1\sim N}$ and $B_{1\sim N}$) in a primitive cell. The C-C interactions are $\alpha_{0}$=2.569 eV, $\alpha_{1}$=0.361 eV, $\alpha_{3}$=-0.032 eV, and $\alpha_{2}$=0.013 eV,$^{13}$ which respectively represent the atomic hopping integral between the nearest-neighbor atoms on the same layer, the atoms \emph{A} or \emph{B} from the nearest-neighbor
layers, the atoms \emph{A} and \emph{B} from two nearest-neighbor layers, and the atoms \emph{A} or \emph{B} from two next
nearest-neighbor layers. The first Brillouin zone
with symmetric points $\Gamma$(0,0), M(2$\pi$/(3\emph{b}),0) and
K(2$\pi$/(3\emph{b}),2$\pi$/(3$\sqrt{3}$\emph{b})) is the same as
that of a graphene sheet.

The matrix representation is a $2\emph{N}\times2\emph{N}$ Hermitian
matrix consisting of 2$\times$2 blocks $h_{l,l}$, $h_{l,l+1}$,
$h_{l+1,l}$, $h_{l,l+2}$, $h_{l+2,l}$ and other zero blocks, where
\emph{l} stands for the \emph{l}th sheet and \emph{l}$\leq$\emph{N}.
The diagonal block $h_{l,l}$ describes the intralayer interactions, and the off-diagonal blocks
$h_{l,l+1}=h_{l+1,l}$, $h_{l,l+2}=h_{l+2,l}$ come from the interlayer interactions. An electric field perpendicular to the graphene sheets, adds an electric potential $\emph{U}_{l}$ to the site energy of the $l$th-layer carbon atoms. They are expressed by

\begin{equation}
h_{l,l}=\left(
          \begin{array}{cc}
            \emph{U}_{l} & \alpha_{0}f(\textbf{k}) \\
            \alpha_{0}f^{*}(\textbf{k}) & \emph{U}_{l} \\
          \end{array}
        \right),\notag
\end{equation}

\begin{equation}
h_{l,l+1}=h_{l+1,l}=\left(
                      \begin{array}{cc}
                        \alpha_{1} & \alpha_{3}f(\textbf{k}) \\
                        \alpha_{3}f^{*}(\textbf{k}) & \alpha_{1} \\
                      \end{array}
                    \right),~and
\end{equation}

\begin{equation}
h_{l,l+2}=h_{l+2,l}=\left(
                      \begin{array}{cc}
                        \alpha_{2} & 0 \\
                        0 & \alpha_{2} \\
                      \end{array}
                    \right).\notag
\end{equation}
$\emph{f}(\textbf{k})=\sum_{j=1}^{3}\textrm{exp(}i\textbf{k}
\cdot\mathbf{r_{j}})$, \textbf{k} is the wave vector, and $\mathbf{r_{j}}$ is the position vector of the nearest-neighbor atom \emph{B} (\emph{A}) with respect to the atom \emph{A} (\emph{B}) on
the same layer. The set of the electrostatic potentials for a general number of layers can be derived by a self-consistent calculation.$^{33}$ In few-layer graphenes ($N=2\sim 4$), the potential difference between layers is roughly linear, i.e., $\emph{U}_{l}=(-\emph{N}+2\emph{l}-1)\emph{eFc}/2$, where \emph{e} is the electric charge and $F$ is the strength of the effective field due to the screening effect. Through the diagonalization of Hamiltonian matrix, the energy dispersions
$E_{n}^{c,v}(\textbf{k})$ and the wave functions
$\Psi_{n}^{c,v}(\textbf{k})$ are obtained, where \emph{n} is the
band index, and \emph{c} and \emph{v} respectively denotes the
unoccupied and the occupied states. The wave function is the linear
superposition of the periodic tight-binding functions
$\psi_{lh}(\textbf{k})$, i.e.,
$\Psi_{n}(\textbf{k})$=$\sum_{lh}u_{nlh}(\textbf{k})\psi_{lh}(\textbf{k})$ with \emph{h} denoting the atom \emph{A} or \emph{B}.

The electron-electron (e-e) Coulomb interactions will induce charge screening and thus dominate excitation properties. In a multilayer graphene, charges in all layers would participate the screening process. Within the random-phase approximation, the $N\times N$ dielectric-function matrix is written as$^{24,26}$
\begin{equation}
\epsilon_{ll'}(\textbf{q},\omega)= \epsilon_{0}\delta_{ll'}
-\sum_{m}V_{lm}^{\textrm{ext}}(q)P_{ml'}^{(1)}(\textbf{q},\omega).
\end{equation}

The momentum transfer \textbf{q} is conserved during the e-e Coulomb interactions. \textbf{q}=(\emph{q}cos$\phi$,\emph{q}sin$\phi$), where $\phi$ is the angle of the transferred momentum with respect to $\Gamma$M. $\epsilon_{0}$=2.4 is the
background dielectric constant. The external Coulomb potentials for interlayer and intralayer are
$V_{ll'}^{\textrm{ext}}(q)=v_{q}e^{-q|l-l'|c}$, where
$v_{q}=2\pi$$e^2/q$. The linear bare response function or the RPA bubble is

\begin{align}
P_{ll'}^{(1)}(\textbf{q},\omega)&=2\sum_{\textbf{k}}
\sum_{nn'}(\sum_{h}u_{nlh}(\textbf{k})u_{n'lh}^{*}(\textbf{k}+\textbf{q}))\notag
\\&\times(\sum_{h'}u_{nl'h'}^{*}(\textbf{k})u_{n'l'h'}(\textbf{k}+\textbf{q}))\notag
\\&\times\frac{f(E_{n}(\textbf{k}))-f(E_{n'}(\textbf{k}+\textbf{q}))}
{E_{n}(\textbf{k})-E_{n'}(\textbf{k}+\textbf{q})+\hbar\omega+i\delta}.
\end{align}
\emph{f}($E_{n}$(\textbf{k}))=1/[1+exp($E_{n}$(\textbf{k})-$\mu$(T))/($k_{B}$T)]
is the Fermi-Dirac distribution, where $\mu$(T) is the
temperature-dependent chemical potential and $k_{B}$ is the Boltzmann
constant. $\delta$ is the energy broadening width due to various deexcitation mechanisms. The factor of 2 accounts for the spin degeneracy, and the valley degeneracy is included by summing over all states in the first Brillouin zone. The interlayer polarizations ($l\neq l'$) vanish when the interlayer atomic interactions are absent.

The loss function, or the inelastic scattering probability, can be obtained through detailed calculations.$^{26}$ It can be written as
\begin{align}
\textrm{Im}[-1/\varepsilon]\equiv\sum_{l}\textrm{Im}[-V^{eff}_{ll}(\mathbf{q},\omega)]/(\sum_{ll'}V_{ll'}^{\textrm{ext}}(q)/\emph{N}),
\end{align}
where the effective Coulomb potential is defined as $[V^{eff}]=[\epsilon]^{-1}[V^{ext}]$. The denominator is the average value of the external potentials on the $N$ layers. The screened response function is used to understand the collective excitations of the low-energy $\pi$ electrons.

\bigskip
\bigskip
\centerline {\textbf {III. RESULTS AND DISCUSSION}}%
\bigskip
\bigskip

We begin our discussion with trilayer AA-stacking graphene. The low-energy bands exhibit three sets of linear bands $\pi_{1}$, $\pi_{2}$, and $\pi_{3}$ from low to high energy, as shown in Fig. 1 by the solid lines in different colors. If the interactions $\alpha_{2}$ and $\alpha_{3}$ are considered small, the energy dispersion of the three pairs of linear bands can be roughly expressed as: $\pi_{1}=(\alpha_{2}-\sqrt{8}\alpha_{1})/2\pm(\alpha_{0}-\sqrt{8}\alpha_{3}/2)|f(\mathbf{k})|$; $\pi_{2}=-\alpha_{2}\pm \alpha_{0}|f(\mathbf{k})|$; $\pi_{3}=(\alpha_{2}+\sqrt{8}\alpha_{1})/2\pm(\alpha_{0}+\sqrt{8}\alpha_{3}/2)|f(\mathbf{k})|$, where $\emph{f}(\textbf{k})=\sum_{j=1}^{3}\textrm{exp(}i\textbf{k}
\cdot\mathbf{r_{j}})$ and $\mathbf{r_{j}}$ is the position vector of the nearest-neighbor B (A) atom measured from A (B) atom, as introduced in Method. Among the three pairs, $\pi_{2}$ band shows features very similar to those of an undoped monolayer graphene. It has an intersection around zero energy with a slope of $\pm3\alpha_{0}b/2$, where $b$ is the atom-atom length. The wavefunction of this band reveals that the charge distributions are spatially located on the outmost layers, $L1$ and $L3$, between which exists only very weak interaction $\alpha_{2}$. With respect to the $\pi_{2}$ band, the $\pi_{1}$ ($\pi_{3}$) band has an energy down-shift (up-shift) with its Fermi momentum moving away from the $K$ point by the magnitude $2\sqrt{2}\alpha_{1}/(3\alpha_{0}\emph{b})$ as indicated by the black arrows. Those subband shifts are mainly dominated by the interaction $\alpha_{1}$. The slopes of both $\pi_{1}$ and $\pi_{3}$ bands are modified by the interaction $\alpha_{3}$, which is the chief cause of the particle-hole asymmetry of the whole system. In the presence of an electric field $F$, the $\pi_{1}$ band gains more free electrons and the $\pi_{3}$ band more holes (the dashed lines) as their Fermi momentum further increases to $k_{F}\approx2[2\alpha_{1}^{2}+(eFc)^{2}]^{1/2}$/(3$\alpha_{0}$\emph{b}) measured from the $K$ point. Accordingly, when $F$ is much greater than $\alpha_{1}$, $k_{F}$ is roughly proportional to $F$. For a strictly linear energy dispersion, that would imply that the carrier density is proportional to $F^{2}$.



The main features of the band structure reflect directly on the bare response functions. Both the intralayer polarizations ($P_{11}^{(1)}=P_{33}^{(1)}$ and $P_{22}^{(1)}$) and the interlayer polarizations  ($P_{12}^{(1)}$=$P_{21}^{(1)}$=$P_{23}^{(1)}$=$P_{32}^{(1)}$ and $P_{13}^{(1)}$=$P_{31}^{(1)}$) contribute to the dynamic charge screening (see methods). The former and the latter respectively describe the charge correlations on same layers and between different layers through the Coulomb interactions. Figure 2 displays the real part (the left panels) and imaginary part (the right panels) of the various polarizations as a function of frequency. The energy broadening width is set to $\delta$=2 meV. The two parts are connected by the Kramers-Kronig relations and exhibit similar peak structures, i.e., the square-root divergences. Each of those peaks represents one major single-particle excitation (SPE) channel and the peaks are distributed over the three regions in energy: (i) $\omega_{SPE}^{1}\approx3/2\alpha_{0}bq$, (ii) $\omega_{SPE}^{2}\approx\sqrt{2\alpha_{1}^{2}}\pm3/2\alpha_{0}bq$, and (iii) $\omega_{SPE}^{3}\approx2\sqrt{2\alpha_{1}^{2}}\pm3/2\alpha_{0}bq$, as indicated by the green dashed lines. The peak intensity, related to the density of electron-hole (e-h) pairs and the charge distributions, is determined by the transition channel and the type of the polarization function. The first group of peaks centered at $\omega_{SPE}^{1}$ results from the charge scattering within the same pair, i.e., $\pi_{1}\rightarrow\pi_{1}$, $\pi_{2}\rightarrow\pi_{2}$, and $\pi_{3}\rightarrow\pi_{3}$. It carries very little spectral weight because of the small density of e-h pairs. The second group of peaks located around $\omega_{SPE}^{2}$ corresponds to transitions between the nearest pairs, $\pi_{1}\rightarrow\pi_{2}$ and $\pi_{2}\rightarrow\pi_{3}$ as illustrated by the purple arrows in Fig. 1. It shows two double peaks positioned at $\sqrt{2\alpha_{1}^{2}}+3/2\alpha_{0}bq$ and $\sqrt{2\alpha_{1}^{2}}-3/2\alpha_{0}bq$. The double peaks mainly come from the disparity of the energy spacing between the $\pi_{1}$ and $\pi_{2}$ bands and the energy spacing between the $\pi_{2}$ and $\pi_{3}$ bands at the K-point. This disparity is caused by the interaction $\alpha_{2}$. This group of peaks is prominent in $P_{11}^{(1)}$ and $P_{13}^{(1)}$, but completely absent in $P_{22}^{(1)}$ and  $P_{12}^{(1)}$. The latter is due to the spatial charge distribution of the $\pi_{2}$ band being confined to $L1$ and $L3$ only as mentioned earlier. The third group of peaks, centered around $\omega_{SPE}^{3}$, is dominated by the transition of $\pi_{1}\rightarrow\pi_{3}$, as illustrated by the brown arrows in Fig. 1. Similar to the second group of peaks, it also consists of two double peaks, but in this case the splitting of the peak is due to the different slopes of $\pi_{1}$ and $\pi_{3}$ bands and the change in the slope of a single linear band below and above the intersection. This group of peaks exists in all types of polarizations since the wavefunctions of the $\pi_{1}$ and $\pi_{3}$ bands spread through all layers spatially. It is noted that extra doping ($E_{F}\neq 0$) would not change the main structures of polarization functions but only the strength of divergent peaks. That is because the interlayer atomic interactions induce a number of free carriers in AA-stacking graphene, the effect similar to doping.

An electric field shifts the linear subbands ($\pi_{1}$ and $\pi_{3}$), increases the density of free carriers, changes the charge distributions, and then alters the peak frequencies and heights in the polarization functions, as shown by the red lines in Fig. 2. $\omega_{SPE}^{1}$ is fixed, while $\omega_{SPE}^{2}$ and $\omega_{SPE}^{3}$ are shifted to higher energies with the displacements proportional to the field strength. The behaviors of the peak heights are more complex. The peak at $\omega_{SPE}^{1}$ is apparently enhanced in $P_{11}^{(1)}$ since extra free electrons (holes) are induced mainly on $L1$ ($L3$). As for the second group of peaks (originating from the transitions $\pi_{1}\rightarrow\pi_{2}$ and $\pi_{2}\rightarrow\pi_{3}$), its peak heights are more dominated by the charge distributions. The wavefunction correlations between $\pi_{1}$ and $\pi_{2}$ ($\pi_{2}$ and $\pi_{3}$) bands are enhanced for the parts involving $L1$ and $L2$ ($L2$ and $L3$) but is lowered for that involving $L3$ ($L1$). Therefore, except for $P_{13}^{(1)}$ the second group of peaks is enhanced in all types of polarizations including its occurrence in $P_{22}^{(1)}$ and $P_{12}^{(1)}$ which are absent at $F$=0. Unlike the former two groups of peaks, the third group of peaks is always weakened owing to the reduced wavefunction correlations between $\pi_{1}$ and $\pi_{3}$ bands. The three groups of SPE peaks will respectively contribute to a prominent peak in loss function discussed later.

The loss function, defined as Im[-1/$\epsilon(\textbf{q},\omega)$],
is useful for understanding the collective excitations that can be measured in inelastic light scattering and electron scattering spectroscopy. Distinct from the plateau structure in a monolayer graphene (undoped), the trilayer graphene gives rise to three prominent peaks at low frequency, as shown in Fig. 3 by the blue line. The first and the highest peak is dominated by the intrapair band transitions, especially the $\pi_{1}$ and $\pi_{3}$ bands that have relatively more free charges compared to the $\pi_{2}$ band. From the momentum dependence of the frequency, this excitation mode is known as a two-dimensional (2D) acoustic plasmon (AP) with the charges oscillating in phase and the energy vanishing at zero momentum. The second peak is closely related to the second group of peaks in $P_{11}^{(1)}$ and $P_{13}^{(1)}$, which implies that the plasmon charge oscillations occur on $L1$ and $L3$ only. As for the third peak, it can be associated with the third group of peaks existing in all types of polarization functions, which means that the collective oscillations occur on all layers. The two plasmons have finite energies at zero momentum. They are categorized to 3D optical plasmons (OPs) and suffer a quite Landau damping.


An electric field moves the three plasmon peaks to higher energies and changes their heights. We consider first the case $F$=0.2 (V/{\AA}) and $q$=10 ($10^{5}/cm$) as indicated by the red line in Fig. 3. The intensity of the first peak (the 2D plasmon peak), strongly dominated by the density of free carriers, is largely enhanced. The second plasmon peak is also enhanced because the occurrence of the second group of peaks in $P_{22}^{(1)}$ and $P_{12}^{(1)}$ (Figs. 2(b) and 2(c)) brings about the extra charge oscillations on $L2$. Oppositely, the third peak is weakened due to the descent of the third group of peaks in polarizations. In addition to affecting the original three peaks, the electric field also induces an extra peak (marked by the red star) to the left of the second peak. The generation of this new peak is the counterpart of the peak splitting of the second group of peaks in polarizations as discussed before, and the field enhances the splitting effects on the plasmon excitations. By changing the momentum to a proper range ($40<q<70$), another peak could become visible below the first major peak, as shown by the green line below a green star. Its appearance might lower the threshold excitation frequency and could be important in real transport applications.


The transferred momentum dependence of both the plasmon frequency and intensity are shown in Fig. 4. The strong dependence of $\omega_{P}$ on $q$ indicates that the electron wave can propagate with a high group velocity $d\omega_{P}(q)/dq$. At $F=0$ there exist three plasmon modes (the blue curves). The lowest one belongs to the acoustic mode and its frequency at small $q$ can be well fitted by the relation $\omega_{p}\sim\sqrt{q}$. The $q$-dependent behavior is similar to that of a 2D electron gas. Its spectral weight gradually decreases with increasing momentum, and quickly damps out after entering the interpair band excitation region (the boundary of which is plotted as the solid line). The two other modes near $q=0$ have frequencies slightly higher than the corresponding SPE energies of about $\sqrt{2}\alpha_{1}$ and $2\sqrt{2}\alpha_{1}$. They belong to OPs with the frequency strongly affected by the interaction $\alpha_{1}$ and with a linear momentum dispersion at high $q$.



The electric field strongly modifies the AP dispersion, increases the OP frequencies, and creates new branches. It enhances the $q$ dependence of the acoustic plasmon as well as its intensity, and prolongs its lifetime. The effects are caused by two reasons. One is the increase of carrier density from the down-shift or up-shift of the linear subbands. The other is that the gap between the SPE boundaries of the (main) intrapair and the interpair band transitions is widened. Therefore, the acoustic plasmon can exist in a larger region without suffering a Landau damping from the interpair band excitations. The two new plasmon branches created by the electric field are indicated by the red stars. The lower one disperses along the SPE boundary of the (main) intrapair band transitions. Its intensity reaches the maximum right before it enters the region of the nearest-interpair band excitations (the boundary is plotted in dashed line). The higher one is adjacent to the major OP branch (at $\omega_{p}\simeq 1.15$ eV when $q$=0). At q=10 points, the two close OP modes correspond to the double peaks in the loss function (the red line in Fig. 3). The branch splitting may be visible before it enters the next-nearest-interpair band SPE region (the dashed line).

The dependence of the plasmon frequency on the field strength (or the carrier density) deserves a further discussion. Figure 5(a) displays the major AP frequency for various numbers of layers. It shows that at low $F$ the AP frequency of bilayer and trilayer graphenes can be approximately fitted by the relation $\omega_{p}\propto N_{d}^{1/4}$, where $N_{d}$ is the electron (hole) density of $\pi_{1}$ ($\pi_{3}$) band. This is a consequence of the linear density of states similar to monolayer graphene.$^{34}$ However, as $F$ keeps increasing, the Fermi momentum moves away from the $K$ point, and the surrounding energy dispersion gradually becomes anisotropic and nonlinear, which causes $\omega_{p}$ to depart from the quartic-root relationship. This implies that there should exist a critical density or a limit of field strength, beyond which the linear energy approximation$^{20,35}$ is no longer applicable. For layer number $N=4$, the dependence of $\omega_{p}$ on $F$ cannot be described by a simple equation because the low energy bands are apparently distorted. In other words, it is always necessary to consider the exact $\pi$-band structure. Additionally, the field-induced APs are shown in Fig. 5(b). Their frequencies depend on $F$ in a roughly linear relationship. To induce the excitation mode the field strength must exceed a critical value, which depends on the number of layers and the magnitude of the transferred momentum.


The dependence of the OPs on the field strength in terms of their frequency and intensity is shown in Figs. 6(a)-(c), respectively for layer number $N$=2, 3, and 4. The amount of OPs at zero field is equal to $N-1$, a feature that could be used to identify the number of layers. The OP frequencies, reflecting the energy differences between different $K$-point states in linear bands (Fig. 1), increase monotonically with $F$. Whether the spectral weight is enhanced or reduced depends on the transition channel. The lowest OP dominated by the nearest-interpair band transitions is largely enhanced with its frequency gradually turning away from the SPE energy (the black dashed lines). Other higher OPs, in contrast, are quickly weakened with their frequencies approaching the SPE energies. The extra branches indicated by the green arrows can only occur for the cases of $N>2$ and when the field strength is sufficiently strong. They carry strong spectral weights for low energies and larger number of layers. Their appearance stems from the nonuniform subband slopes and energy spacings between the different $K$-states, which are the results of considering all the significant interlayer atomic interactions and the exact $\pi$-band structures. With the increase of the number of layers, the dispersions of the field-induced branches become more complex since more pairs of linear subbands are involved. However, in general, it may be observed that the extra branches approach the major ones (those exist at $F=0$) as $F$ keeps on increasing, because the uniformity of the subband slopes and the energy spacings of the $K$-states is improved.

A comparison with different stacking sequences is made. Figure 7 shows the loss spectra of trilayer ABA- and ABC-stacking graphenes with the corresponding energy bands depicted in the insets. The low-frequency plasmons hardly exist due to the insufficient free carriers (the green lines). However, an electric field can oscillate the energy bands, even cause a prominent band gap in ABC-stacking graphen, create extra band-edge states, and lead to noticeable interband plasmons (the red lines). They belong to optical modes and always suffer Landau damping. Similar to AA-stacking graphene, the OP frequencies always increase with rising $F$, while their intensities can be enhanced or weakened depending on the transition channel. The absence of acoustic plasmons is a prominent characteristic that differs from AA-stacking graphene. Accordingly, it is known that the low-energy physical properties and their responses to external fields are strongly dominated by the interlayer atomic interactions and the stacking structures.

\bigskip
\bigskip
\centerline {\textbf {IV. SUMMARY AND CONCLUSIONS}}%
\bigskip
\bigskip

In this work, we have discussed the excitation properties of AA-stacking MLG under a perpendicular electric field. The low band structure of the material is like a combination of $N$ (the number of layers) pairs of linear bands. The electric field shifts the linear subbands, increases the density of free carriers, and alters the charge distributions, as a result, changes the response-function matrix significantly. The energy-loss function has been used to study the collective excitations. In the absence of the field, it displays one AP and $N-1$ OPs at low frequency. The electric field increases the plasmon frequencies, alters their energy spacings, and rearranges their spectral weights. It also induces extra plasmon modes, including a few of OP modes and an acoustic one in a proper momentum range; the latter lowers the threshold excitation frequency. Those significant changes imply that the plasmon effects in MLG could be electrically tunable, a feature that could be useful in electronic applications and might attract more theoretical and experimental researches to this area.

\bigskip

\bigskip

\centerline {\textbf {ACKNOWLEDGMENT}}%

\bigskip

\bigskip

\noindent \textit{Acknowledgments.} The authors thank W. P. Su for helpful discussions. This work was supported by the NSC of Taiwan, under Grant No. 98-2112-M-006-013-MY4.

\newpage

\par\noindent ~~~~$^\star$e-mail address: yarst5@gmail.com

\begin{itemize}


\item[$^{1.}$] C. L. Lu, C. P. Chang, Y. C. Huang, J. M. Lu, C. C. Hwang, and M. F. Lin, J. Phys.: Condens. Matter \textbf{18}, 5849 (2006).

\item[$^{2}$] S. Latil and L. Henrard, Phys. Rev. Lett. \textbf{97}, 036803 (2006).

\item[$^{3}$] K. F. Mak, J. Shan, and T. F. Heinz, Phys. Rev. Lett. \textbf{104}, 176404 (2010).

\item[$^{4}$] F. Zhang, B. Sahu, H. Min, and A. H. MacDonald, Phys. Rev. B \textbf{82}, 035409 (2010).

\item[$^{5}$] W. Bao, L. Jing,  J. Velasco Jr, Y. Lee, G. Liu, D. Tran, B. Standley, M. Aykol, S. B. Cronin, D. Smirnov, M. Koshino, E. McCann, M. Bockrath, and C. N. Lau, Nature Phys. \textbf{7}, 948 (2011).


\item[$^{6}$] E. V. Castro, K. S. Novoselov, S. V. Morozov, N. M. R. Peres, J. M. B. Lopes dos Santos, J. Nilsson, F. Guinea, A. K. Geim, and A. H. Castro Neto, Phys. Rev. Lett. \textbf{99}, 216802 (2007).

\item[$^{7}$] S. Bala Kumar and J. Guo, Appl. Phys. Lett \textbf{98}, 222101 (2011).

\item[$^{8}$] K. Tang, R. Qin, J. Zhou, H. Qu, J. Zheng, R. Fei, H. Li, Q. Zheng, Z. Gao, and J. Lu, J. Phys. Chem. C \textbf{115}, 9458 (2011).

\item[$^{9}$] W. Zhang, C. T. Lin, K. K. Liu, T. Tite, C. Y. Su, C. H. Chang, Y. H. Lee, C. W. Chu, K. H. Wei, J. L. Kuo, and L. J. Li, ACS Nano \textbf{5} 7517 (2011).


\item[$^{10}$] C. H. Lui, Z. Li, K. F. Mak, E. Cappelluti, and T. F. Heinz, Nature Phys. \textbf{7}, 944 (2011).

\item[$^{11}$] T. Khodkov, F. Withers, D. C. Hudson, M. F. Craciun, and S. Russo, Appl. Phys. Lett. \textbf{100}, 013114 (2012).


\item[$^{12}$] J. -C. Charlier, X. Gonze, and J. -P. Michenaud, Carbon \textbf{32}, 289 (1994).


\item[$^{13}$] J. -C. Charlier, and J. -P. Michenaud, Phys. Rev. B \textbf{46}, 4531 (1992).


\item[$^{14}$] Y. H. Ho, Y. H. Chiu, D. H. Lin, C. P. Chang, and M. F. Lin, ACS Nano \textbf{4}, 1465 (2010).

\item[$^{15}$] X. F. Wang, and T. Chakraborty, Phys. Rev. B \textbf{75}, 041404 (2007).

\item[$^{16}$] G. Borghi, M. Polini, R. Asgari, and A. H. MacDonald, Phys. Rev. B \textbf{80}, 241402 (2009).


\item[$^{17}$] J. K. Lee, S. C. Lee, J. P. Ahn, S. C. Kim, J. I. B. Wilson, and P. John, J. Chem. Phys. \textbf{129}, 234709 (2008).


\item[$^{18}$] Z. Liu, K. Suenaga, P. J. F. Harris, and S. Iijima, Phys. Rev. Lett. \textbf{102}, 015501 (2009).


\item[$^{19}$] I. Lobato and B. Partoens, Phys. Rev. B \textbf{83}, 165429 (2011).

\item[$^{20}$] Y. F. Hsu and G. Y. Guo, Phys. Rev. B \textbf{82}, 165404 (2010).


\item[$^{21}$] Y. Xu, X. Li, and J. Dong, Nanotech. \textbf{21}, 065711 (2010).

\item[$^{22}$] Y. H. Ho, J. Y. Wu, R. B. Chen, Y. H. Chiu, and M. F. Lin, Appl. Phys. Lett. \textbf{97}, 101905 (2010).


\item[$^{23}$] P. R. Wallace, Phys. Rev. \textbf{1947}, 71, 622-634 (1947).


\item[$^{24}$] J. H. Ho, C. L. Lu, C. C. Hwang, C. P. Chang, and M. F. Lin, Phys. Rev. B \textbf{74}, 085406 (2006).


\item[$^{25}$] H. Ehrenreich, and M. H. Cohen, Phys. Rev. \textbf{115}, 786 (1959).


\item[$^{26}$] J. H. Ho, C. P. Chang, and M. F. Lin, Phys. Lett. A \textbf{352}, 446 (2006).


\item[$^{27}$] K. F. Mak, C. H. Lui, J. Shan, and T. F. Heinz, Phys. Rev. Lett. \textbf{102}, 256405 (2009).

\item[$^{28}$] J. P. Reed, B. Uchoa, Y. I. Joe, Y. Gan, D. Casa, E. Fradkin, and P. Abbamonte, Science \textbf{330}, 805 (2010).

\item[$^{29}$] R. Hambach, C. Giorgetti, N. Hiraoka, Y. Q. Cai, F. Sottile, A. G. Marinopoulos, F. Bechstedt, and L. Reining, Phys. Rev. Lett. \textbf{101}, 266406 (2008).


\item[$^{30}$] S. Y. Shin, C. G. Hwang, S. J. Sung, H. S. Kim, and J. W. Chung, Phys. Rev. B \textbf{83}, 161403 (2011).

\item[$^{31}$] V. B. Jovanovi\'{c}, I. Radovi\'{c}, D. Borka, and Z. L. Mi\v{s}kovi\'{c}, Phys. Rev. B \textbf{84}, 155416 (2011).

\item[$^{32}$] T. Eberlein, U. Bangert, R. R. Nair, R. Jones, M. Gass, A. L. Bleloch, K. S. Novoselov, A. Geim, and P. R. Briddon, Phys. Rev. B \textbf{77}, 233406 (2008).





\item[$^{33}$] M. Koshino, Phys. Rev. B \textbf{81}, 125304 (2010).


\item[$^{34}$] A. H. Castro Neto, F. Guinea, N. M. R. Peres, K. S. Novoselov, and A. K. Geim, Rev. Mod. Phys. \textbf{81}, 109 (2009).


\item[$^{35}$] J. W. Gonz$\acute{a}$lez, H. Santos, M. Pacheco, L. Chico, and L. Brey, Phys. Rev. B \textbf{81}, 195406 (2010).

\end{itemize}

\bigskip \vskip0.6 truecm

\noindent

\newpage

\centerline {\Large \textbf {Figure Captions}}

\vskip0.3 truecm

Figure 1. The energy bands of trilayer AA-stacking graphene at zero (solid lines) and external field $F$=0.2 V/{\AA} (dashed lines).

\vskip0.5 truecm

Figure 2. The real and imaginary parts of the intralayer ($l=l'$) and interlayer ($l\neq l'$) response functions $P_{ll'}^{(1)}$ in zero (blue lines) and external field $F$=0.2 V/{\AA} (red lines) for $q$=10 in the unit of $10^{5}$ $cm^{-1}$.
\vskip0.5 truecm

Figure 3. The energy loss functions at $F=0$ (the blue line) and $F$=0.2 V/{\AA} (the red line) for $q$=10. That for $q=50$ is shown by the green line.

\vskip0.5 truecm

Figure 4. The momentum-dependent frequencies and intensities of the low-frequency plasmons at $F=0$ (the blue curves) and  $F$=0.2 V/{\AA} (the red curves). The black solid and dashed lines respectively represent the SPE boundaries at the two conditions. The main SPE regions at $F=0$ are named according to the most contributing transition channels.

\vskip0.5 truecm

Figure 5. The field-strength-dependent frequencies of (a) the major APs, and (b) the field-induced APs. The curves from low to high represent the different numbers of layers $N$=2, 3, and 4, respectively. The dashed lines are the $\omega_{p}\propto N_{d}^{1/4}$ relationships, where $N_{d}$ is the carrier density.

\vskip0.5 truecm

Figure 6. The field-strength-dependent frequencies and intensities of OPs at (a) $N$=2, (b) $N$=3, and (c) $N$=4. The SPE frequencies are plotted by the black dashed lines. The green arrows indicate the field-induced OPs.

\vskip0.5 truecm

Figure 7. The loss spectra of (a) ABA and (b) ABC graphene at $F=0$ (the green line) and $F$=0.1 V/{\AA} (the red line). The corresponding energy bands are depicted in the insets.

\newpage

\begin{figure}
\center
\includegraphics[width=10cm]{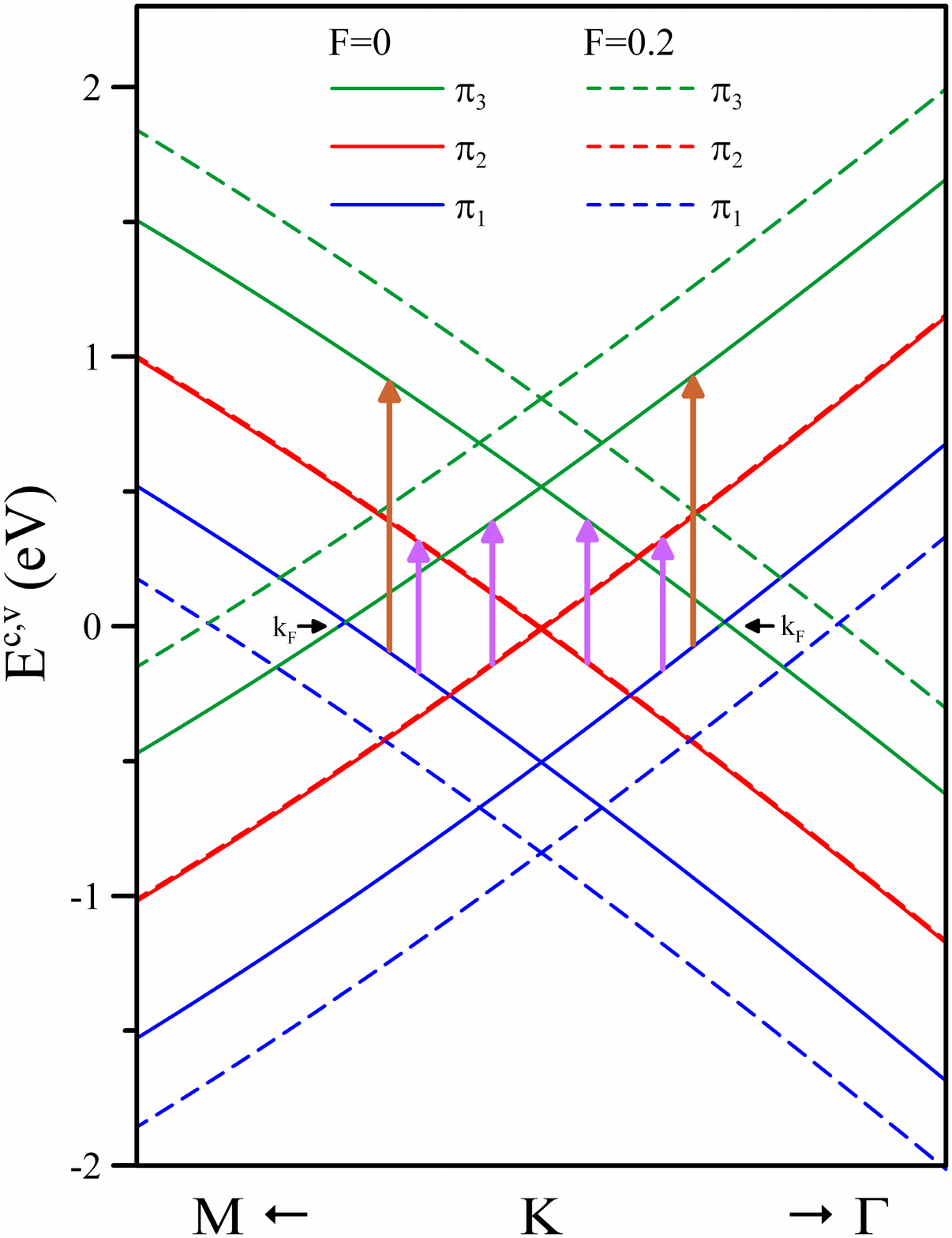}
\end{figure}

\newpage

\begin{figure}
\center
\includegraphics[width=10cm]{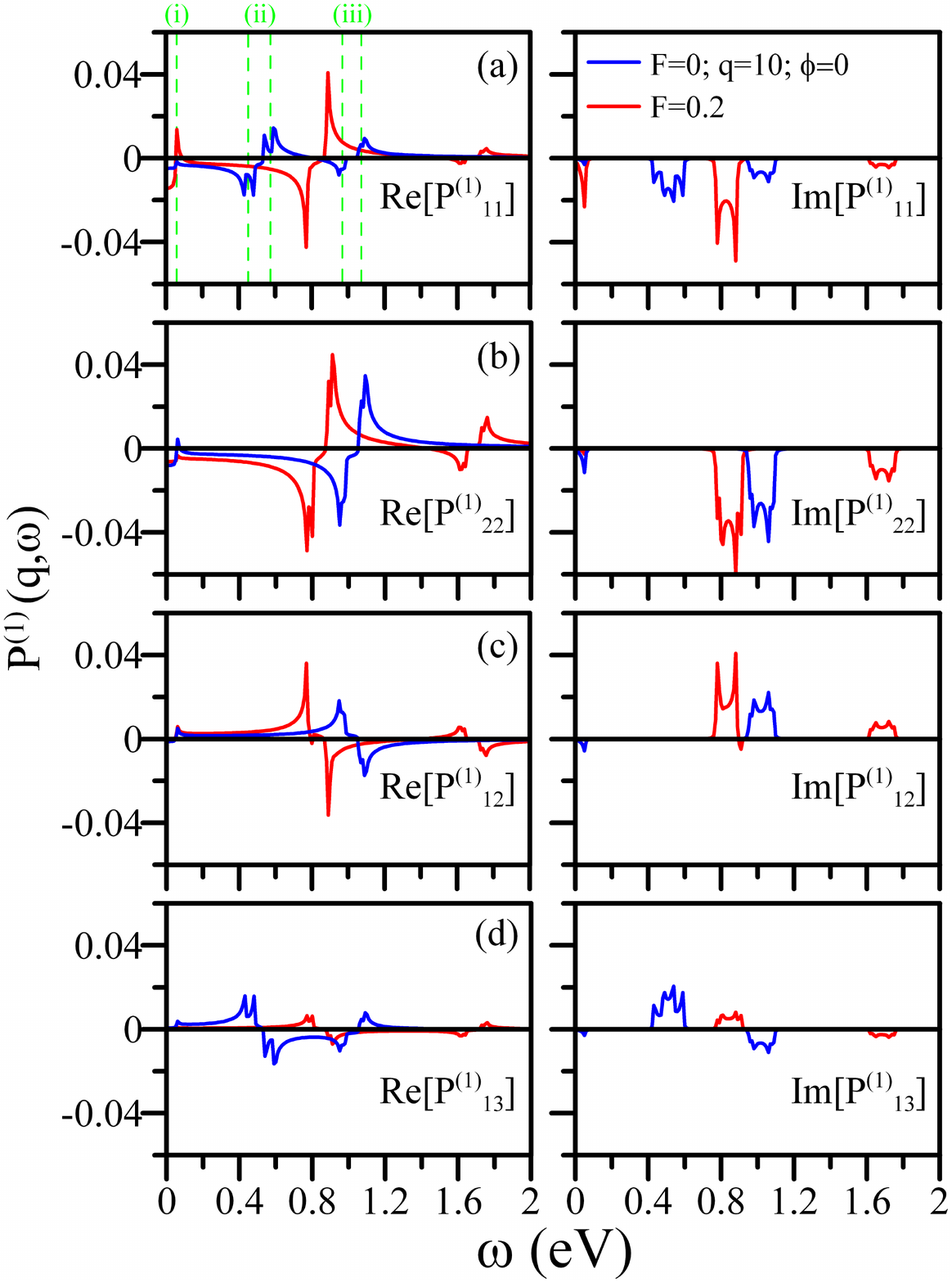}
\end{figure}

\newpage

\begin{figure}
\center
\includegraphics[height=1.0\textheight]{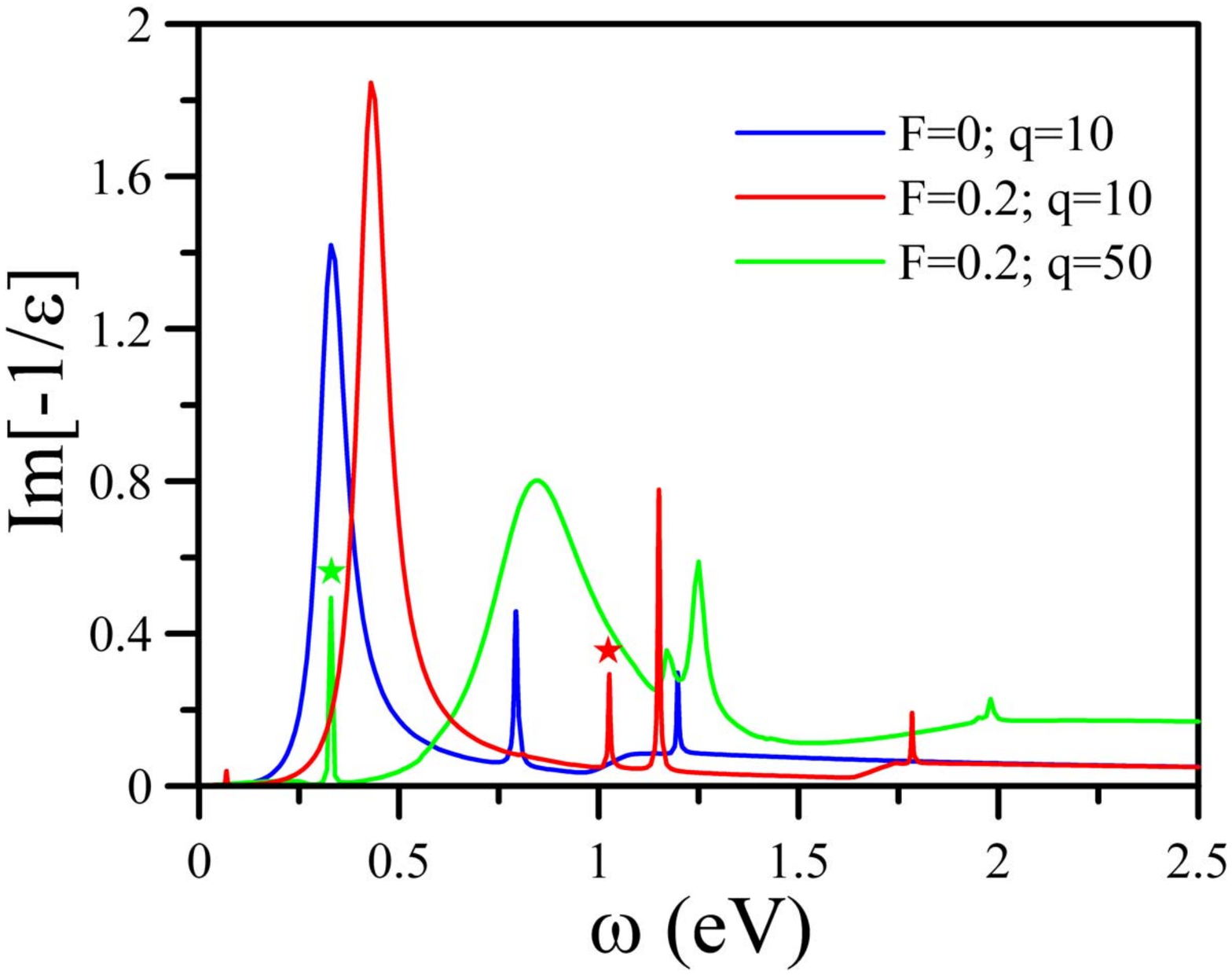}
\end{figure}

\newpage

\begin{figure}
\center
\includegraphics[width=14cm]{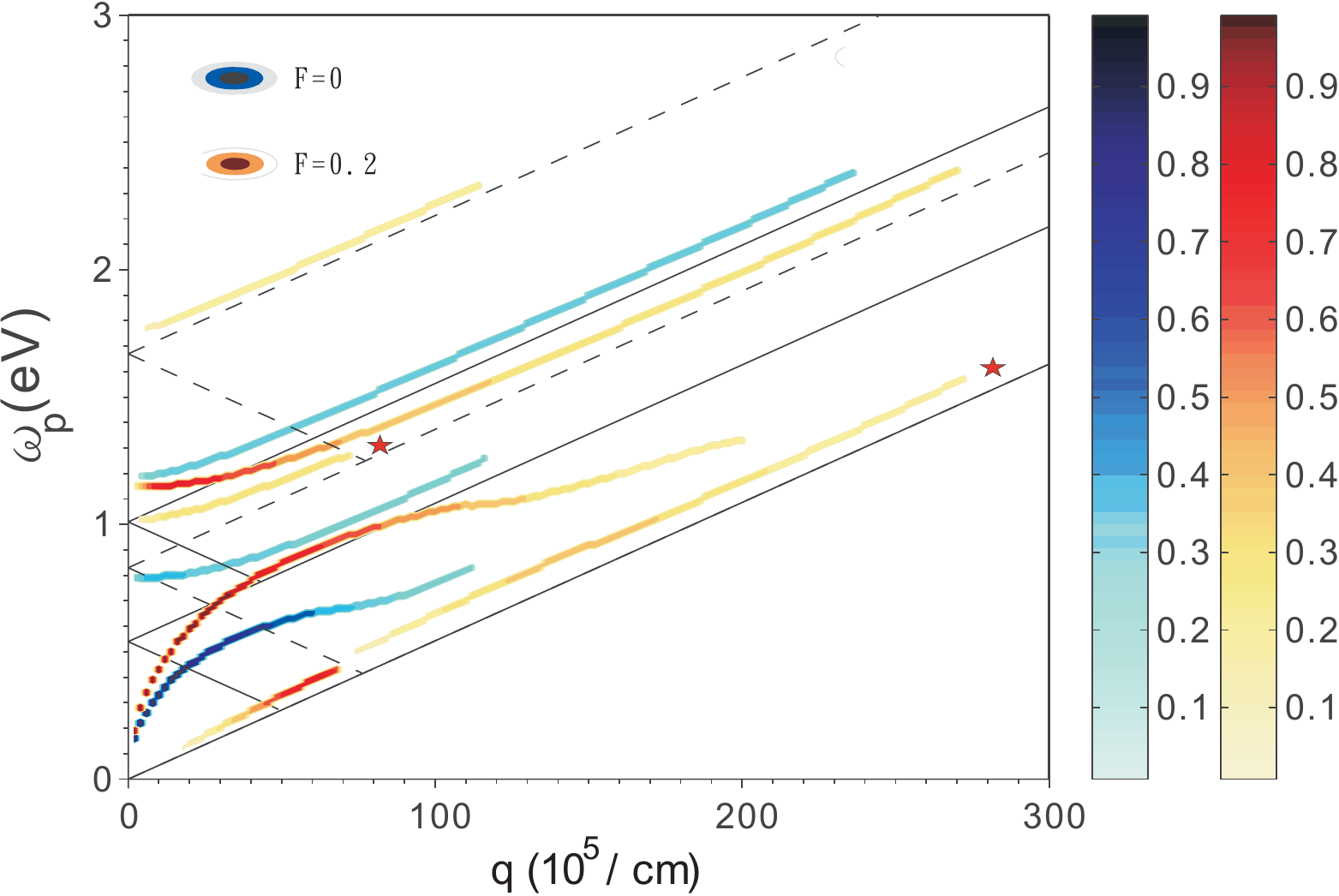}
\end{figure}

\newpage

\begin{figure}
\center
\includegraphics[width=14cm]{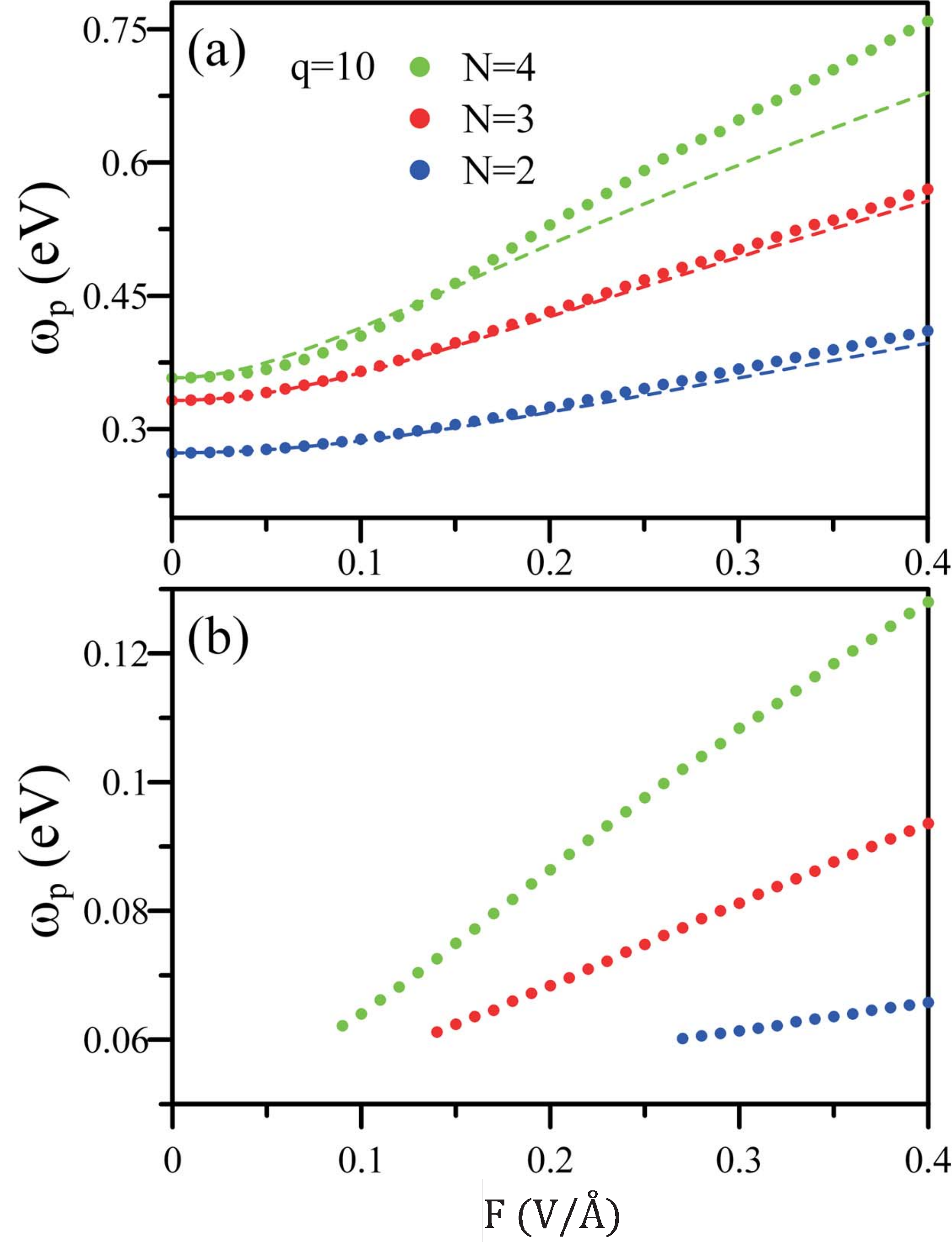}
\end{figure}

\newpage

\begin{figure}
\center
\includegraphics[width=10cm]{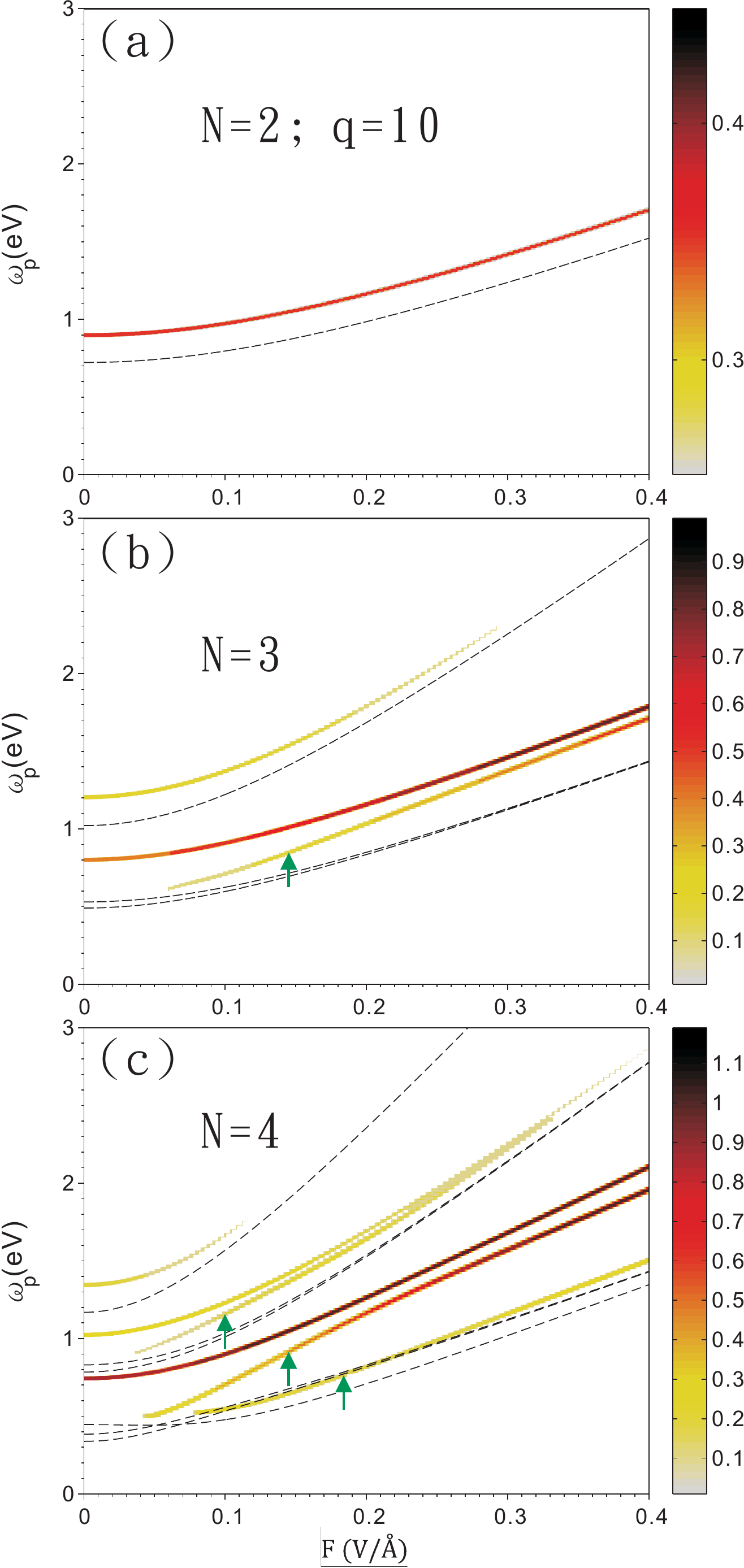}
\end{figure}

\newpage

\begin{figure}
\center
\includegraphics[width=10cm]{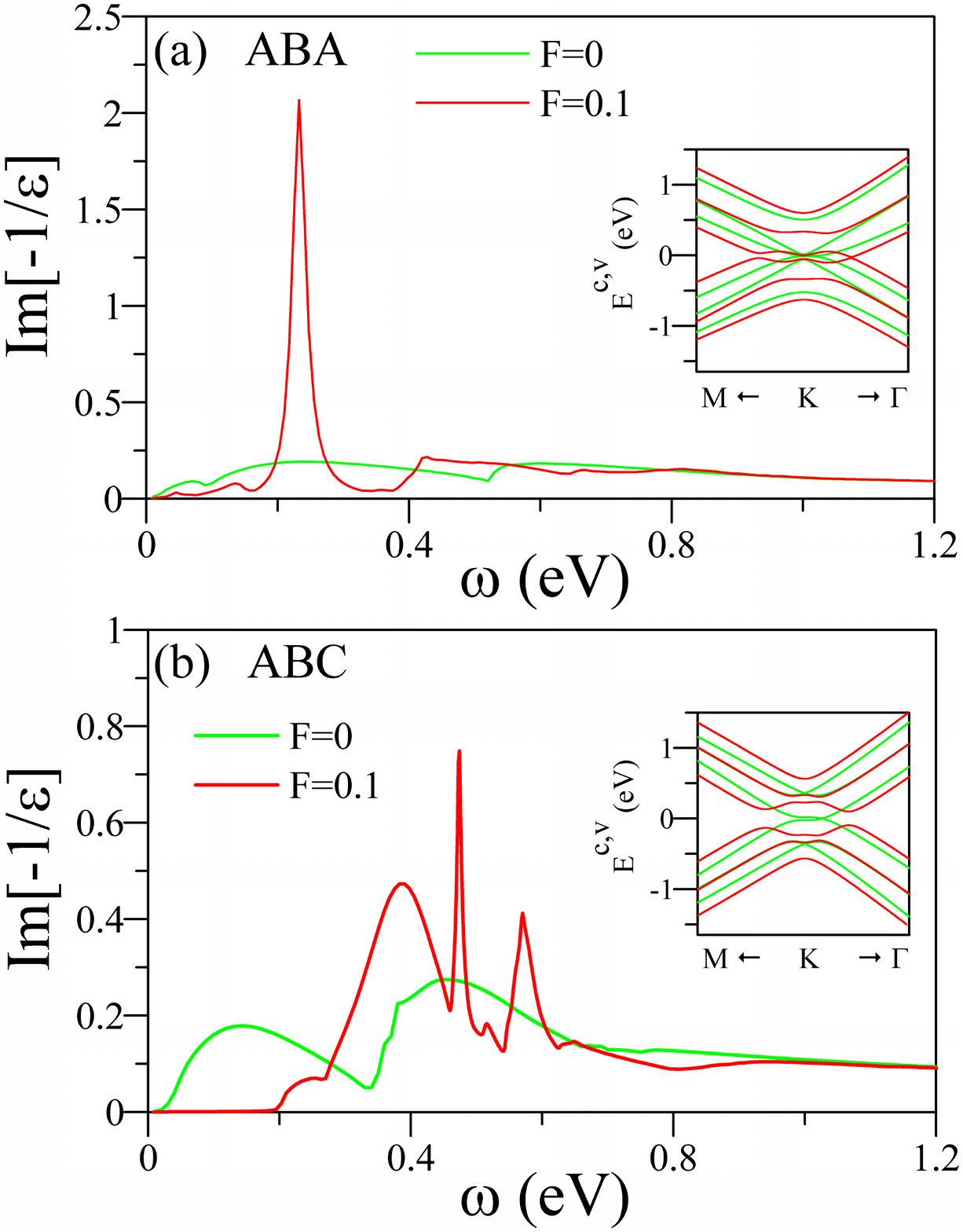}
\end{figure}

\end{document}